# Transplantation of Data Mining Algorithms to Cloud Computing Platform when Dealing Big Data

Yong Wang
College of Engineering & Information Technology
University of Chinese Academy of Sciences
Beijing, China
wangyong@ucas.ac.cn

Zhao Ya-Wei
College of Engineering & Information Technology
University of Chinese Academy of Sciences
Beijing, China
zhaoyw@ucas.ac.cn

*Abstract*—This paper made a short review of Cloud Computing and Big Data, and discussed the portability of general data mining algorithms to Cloud Computing platform. It revealed the Cloud Computing platform based on Map-Reduce cannot solve all the Big Data and data mining problems. Transplanting the general data mining algorithms to the real-time Cloud Computing platform will be one of the research focuses in Cloud Computing and Big Data.

## I. INTRODUCTION

The earliest reference to the term "Big Data" can be traced back to the open source project Nutch of Apache [1]. At that time, Big Data is described as a large amount of batch processing or analyzing data sets used for updating network search index. But as Google Map-Reduce and Google File System (GFS) released, Big Data is not only used to describe a large number of data, but also refers to the speed of data processing [2].

Nowadays, Big Data means to deal with massive complex data. The massive complex data has two key points, one is massive, and the other is complex. "Massive" means that the amount of data is large, and there is more and more real-time processing data. The most important part of the enterprise cost is the time cost, so once the time used in processing massive data is not sustainable, the decision of enterprises will lag market changes. "Complex" means that the data is multiple, and it is no longer just a structured data as past. So the former data models and analyzing theories do not work. It must set up a set of effective analysis theories and models adapt to multivariate data [3]. With the growth in demand of analyzing such data, it has prompted the change from routine analysis to depth analysis, and data analysis has become the essential support for enterprise profits. The modern enterprises are no longer satisfied with the existing data analysis and detection results, but they expect more analysis and forecast on future trends for enhancing their competitiveness. The development of data analysis meets this demand. It gradually evolved from the early stage of statistical analysis to the current age of algorithm optimization (shown in Fig. 1).

The rise of Cloud Computing provides a new solution to Big Data analysis, and many researchers have done extensively studies in the field of Cloud Computing [4-5]. With the analysis of different features of data, Data Mining finds the internal relations in data, as are applicable to discover hidden patterns. But the general data mining algorithms cannot be applied directly to Cloud Computing platform. In the transplantation process to the Cloud Computing platform, the general data mining algorithms need to be reformed. The second part of this paper focused on the technical characteristics of Cloud Computing and expounded Big Data solutions based on it; the third part discussed the transplantation of general data mining algorithms to Cloud Computing platform, with an emphasis on analysis of the transplantation strategy and portability; the fourth part made a summary of this paper, and the future developing directions were discussed.

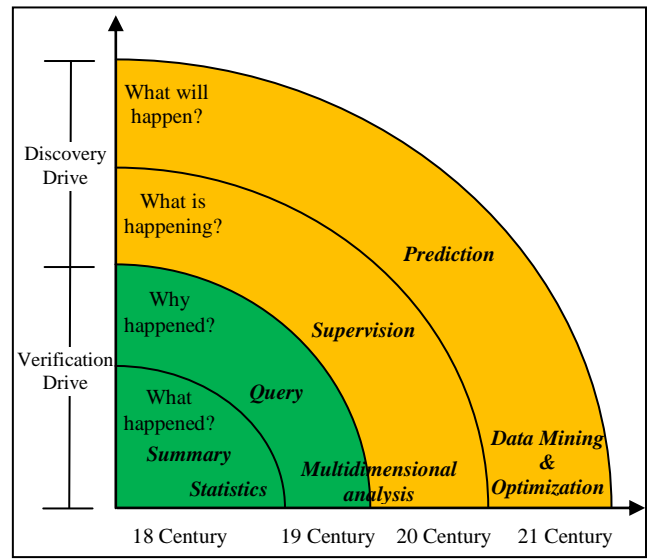

Figure 1. Trends of Data Analysis.

## II. CLOUD COMPUTING FOR BIG DATA

Cloud Computing and Big Data are the both sides of a coin [6]. Cloud Computing is the IT base of Big Data, and Big Data is one of the important applications of Cloud Computing. The 2013 Hype Cycle for Emerging Technologies' maturity curve released by the well-known consulting company Gartner shows that [7], Big Data technology is in the high-speed rising state of transition from embryo stage to inflated expectation stage. Its attention will continue to heat up and make a breakthrough on the technology. Meanwhile Cloud Computing technology is gradually moving towards disillusionment stage, as marks Cloud Computing technology is changing from the concept of speculation to the real mature (shown in Fig. 2). Cloud Computing technology is increasingly mature, and has begun to be widely used.

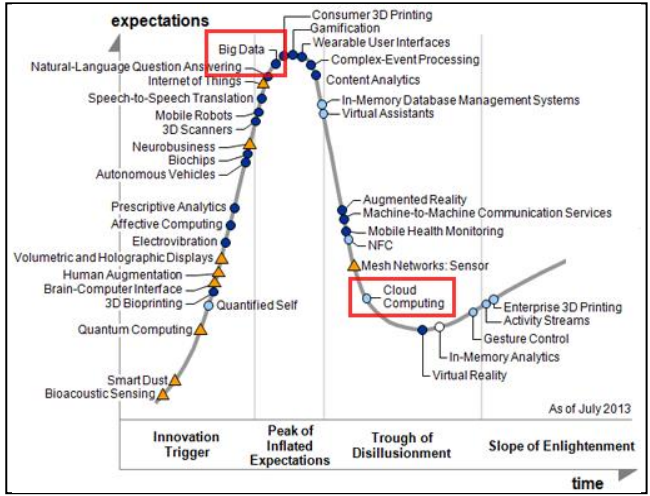

Figure 2. 2013 Hype Cycle for Emerging Technologies.

From 1960s' when John McCarthy envisions that "computation" might someday be organized as a public utility to 2010s' when Cloud 2.0 model is emerging [8], Cloud Computing has undergone multiple developing stages, such as Distributed Computing, Parallel Computing, Cluster Computing, Grid Computing, and now it has developed to a mature level [9].

Figure 3 gives an overview of the relationship between Cloud Computing and other domains that it overlaps with. Web 2.0 covers almost the whole spectrum of service-oriented applications, where Cloud Computing lies at the large-scale side. Supercomputing and Cluster Computing have been more focused on traditional non-service applications. Grid Computing overlaps with all these fields where it is generally considered of lesser scale than supercomputers and Cloud Computing [10].

In the age of Big Data, more data will store in the data center and become the new core assets of enterprises. Cloud Computing technology provides a new way to Big Data storage, processing and analysis. Cloud Computing can treat data as a service, and with the analysis and application of Big Data it unearths the effective data set suitable for a particular scene and topic. Through the mining, Cloud Computing gives people more powerful ability of insight into the future.

Key technologies involved in Cloud Computing includes virtual technology, data storage technology, resource management technology, integration technology, automation technology, and programming model technology [11], and the key of transplantation of general data mining algorithms adapt to the Cloud Computing platform is reforming general data mining algorithms based on the programming model technology. Nowadays, the programming model used in most of "cloud" plans put forward by IT manufacturers is developed based on the idea of Map-Reduce, as is the sub components of Hadoop.

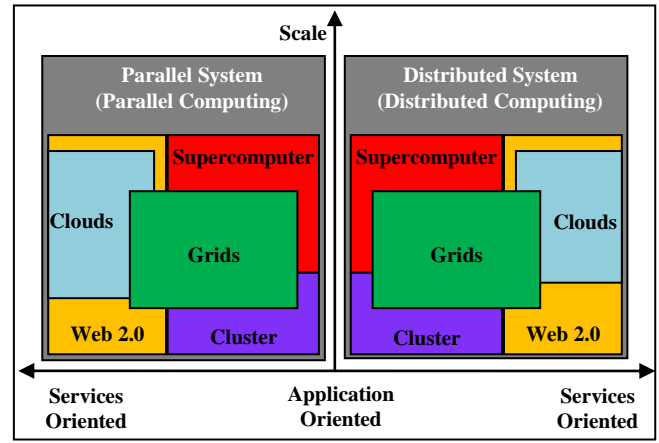

Figure 3. Cloud Computing Overview.

Map-Reduce is a parallel computing framework, and is often used to design the perform layer of Cloud Computing platforms. Although its process of solving problem is as the same as parallel computing (shown in Fig. 4), contrast to the traditional parallel computing program, Map-Reduce provides a simple and powerful interface through the package of parallel processing, fault tolerance, load balance, localization, task scheduling and other details. Using the interface, Map-Reduce can automatically perform the Big Data computing concurrency and distributed execution, which becomes very easy, thus reduces the computational ability of single node. At the same time, it also has good versatility, and a lot of different problems can be solved by Map-Reduce.

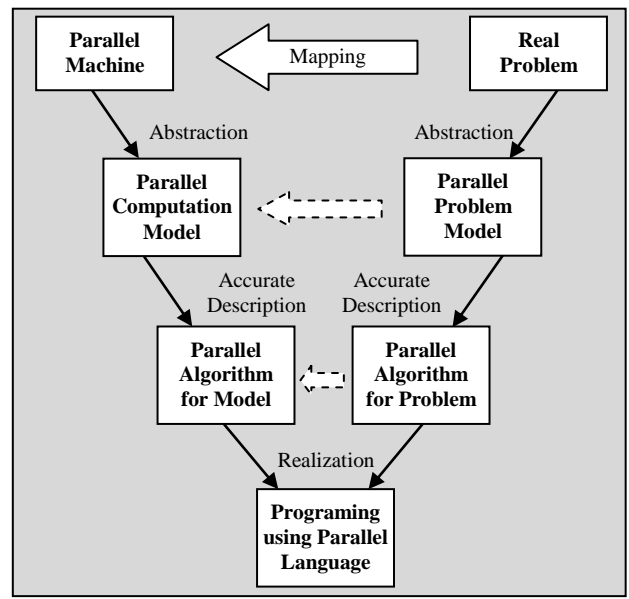

Figure 4. Parallel Computing for Problem Solving.

## III. TRANSPLANTATION OF DATA MINING ALGORITHM

Whether data mining algorithms are able to transplant to the Cloud Computing platform for parallel processing with Map-Reduce or not, it needs specific analysis of specific algorithm [12]. The following are some preliminary and general guiding principles.

- Judge whether the algorithm consists of parallel processing steps: in theoretical analysis it should consider not only the parallelism of the main steps, but also the parallel computing steps of the fine. This is the premise and is also the foundation.

- Judge whether the algorithm meets the Map-Reduce parallel conditions on theory: whether the processing data can be divided by the algorithm and the processing results partitioned can be merged and the final result can be obtained.

- Ensure the time cost of the algorithm with Map-Reduce operations cannot be too much: the algorithm cannot have too many iteration operations, because it is very time-consuming to start Map-Reduce every time. It is necessary to ensure that the algorithm efficiency is promoted after Map-Reduce parallelization; if not there is no need for Map-Reduce parallelization.

- Ensure the correctness of the algorithm after Map-Reduce parallelization: the results from parallel processing should be consistent with the results from serial processing.

TABLE 1 analyzes the efficiency of general data mining algorithms, and summarizes their portability to the framework of Map-Reduce.

## IV. CONCLUSION

The framework of Map-Reduce cannot solve all the Big Data problems, even some problems that can be computed based on parallel processing nodes are not advised to be solved by Map-Reduce parallelization. The main reason is that Map-Reduce framework is based on off-line data processing methods to solve problems, and it is only suitable for the simple computation. Similar frameworks based on off-line data processing methods to solve problems include Pregel (Google proposed, suitable for iterative computation) and Dryad (Microsoft proposed, suitable for complex computation).

The emergence of new application of real-time search, high-frequency trading, social networks promotes the need for real-time data processing. In this Big Data environment, the value of data will decrease over time. The Cloud Computing platform needs a new computing framework based on distributed flow computation, such as Storm (BackType proposed) and S4 (Yahoo proposed). Transplanting the general data mining algorithms to the real-time Cloud Computing platform will be one of the research focuses in Cloud Computing and Big Data.

TABLE I. SUMMARY OF PORTABILITY OF DATA MIINING ALGORITHM

| functionality | Algorithm | Time Complexity | Space Complexity | Map-Reduce |
|---|---|---|---|---|
| Classification | C4.5 | $O(nm^2 + kmn)$ | $O(n)$ | √ |
| | CART | $O(nm^2 + kmn)$ | $O(n)$ | √ |
| | SLIQ | $O(nm^2 + kmn)$ | $O(n)$ | √ |
| | SPRINT | $O(nm^2 + kmn)$ | $O(n)$ | √ |
| | kNN | $O((m+n)d+km)$ | $O(km)$ | √ |
| | Navie Bayes | $O(nkm)$ | $O(|V|km)$ | √ |
| | SVMs | $O(mn^2)$ | $O(lm)$ | |
| | ANNs | $O(nmT)$ | $O(lm)$ | |

$n$ is the number of sample, $m$ is the dimension of sample, $k$ is the value of dimension. Especially, $k$ is the number of neighbor for kNN; $|V|$ is the number of target for Bayes; $l$ is the number of support vector for SVMs; $T$ is the number of iteration and $l$ is the number of hidden neurons for ANNs.

| functionality | Algorithm | Time Complexity | Space Complexity | Map-Reduce |
|---|---|---|---|---|
| Clustering | K-means | $O(nki)$ | $O(k)$ | √ |
| | K-medoids | $O(k(n-k)^2)$ | $O(k)$ | √ |
| | CLARA | $O(ks^2+k(n-k))$ | $O(ks)$ | √ |
| | CLARANS | $O(n^2)$ | $O(ks)$ | √ |
| | DBSCAN | $O(n^2)$ | $O(n)$ | √ |
| | OPTICS | $O(n\log n)$ | $O(n)$ | √ |
| | BIRCH | $O(n)$ | $O(n)$ | |
| | Chameleon | $O(n^2)$ | $O(n)$ | |
| | STING | $O(n)$ | $O(m)$ | |

$n$ is the number of sample, $k$ is the number of cluster, $i$ is the number of iteration, $s$ is the number of sampling, $m$ is the dimension of sample.

| functionality | Algorithm | Time Complexity | Space Complexity | Map-Reduce |
|---|---|---|---|---|
| Association | FP-Tree | $O(nm\log m+2^m)$ | $O(nm+2^m)$ | √ |
| | WFP | $O(nm\log m+2^m)$ | $O(nm+2^m)$ | √ |
| | Apriori | $O(n2^m)$ | $O(2^m)$ | √ |
| | Sampling | $O(n2^m/k)$ | $O(2^m)$ | √ |
| | Partition | $O(n2^m)$ | $O(2^m)$ | √ |
| | DHP | $O(n2^m)$ | $O(2^m)$ | |

$n$ is the number of sample, $m$ is the dimension of sample, $k$ is the ratio of total sample and the sample.

ACKNOWLEDGEMENT

This work is supported by National Natural Science Foundation of China (#61371155).